# Immunogenic cell death triggered by pathogen ligands via host germ line-encoded receptors

Chuang Li[1,2], Yichen Wei[3], Chao Qin[2,4], Shifan Chen[2,5], Xiaolong Shao[6,*]

1 Department of Biological Sciences, Purdue University, West Lafayette, IN, 47906, USA

2 College of Veterinary Medicine, China Agricultural University, Beijing, 100193, China

3 School of Life and Environmental Sciences, The University of Sydney, Sydney, NSW 2006, Australia

4 Section of Infection and Immunity, University of Southern California, Los Angeles, CA, 90089, USA

5 Department of Neuroscience, University of Connecticut of Medicine, Farmington, CT, 06030, USA

6 College of Plant Protection, Nanjing Agricultural University, Nanjing, 210095, China.

Chuang Li: li3258@purdue.edu

Yichen Wei: ywei0330@uni.sydney.edu.au

Chao Qin: chaochin@usc.edu

Shifan Chen: schen@uchc.edu

* Correspondence: Xiaolong Shao: xlshao@njau.edu.cn

**Abstract**

The strategic induction of cell death serves as a crucial immune defense mechanism for the eradication of pathogenic infections within host cells. Investigating the molecular mechanisms underlying immunogenic cell pathways has significantly enhanced our understanding of the host's immunity. This review provides a comprehensive overview of the immunogenic cell death mechanisms triggered by pathogen infections, focusing on the critical role of pattern recognition receptors. In response to infections, host cells dictate a variety of cell death pathways, including apoptosis, pyroptosis, necrosis, and lysosomal cell death, which are essential for amplifying immune responses and controlling pathogen dissemination. Key components of these mechanisms are host cellular receptors that recognize pathogen-associated ligands. These receptors activate downstream signaling cascades, leading to the expression of immunoregulatory genes and the production of antimicrobial cytokines and chemokines. Particularly, the inflammasome, a multi-protein complex, plays a pivotal role in these responses by processing pro-inflammatory cytokines and inducing pyroptotic cell death. Pathogens, in turn, have evolved strategies to manipulate these cell death pathways, either by inhibiting them to facilitate their replication or by triggering them to evade host defenses. This dynamic interplay between host immune mechanisms and pathogen strategies highlights the intricate co-evolution of microbial virulence and host immunity.

**Introduction**

Immunogenic cell death induced by infections caused by pathogens play a major role in host immune responses to eradicate evading bacteria or viruses. Investigation of signaling pathways involved in host innate immunity has revealed the rich and diverse mechanisms that govern the sensing of immune cells to various ligands, particularly pathogen associated molecular patterns (PAMPs)[1]. Upon the recognition of PAMPs, the host germ line-encoded pattern-recognition receptors (PRRs) dictates host antimicrobial responses as well as proinflammatory reactions[2]. Subsequently, PRRs located at the cell surface or intracellularly activate a series of downstream signaling cascades, composed of ligands, receptors, adaptor molecules, kinases and transcription factors[3]. The activation of these signal transduction pathways commands the host to express a wide array of immunoregulatory genes, resulting in the synthesis of cytokines and chemokines that recruits other activated immune cells to eliminate the invading pathogen[4]. While the execution of the innate immune response is accomplished by the actions of phagocytes and antigen-presenting cells, the orchestration of adaptive immunity is facilitated by specialized immune cells. This review mainly discusses pathogen-mediated receptor signaling and cell death in innate immune cells, with an emphasis on macrophages and dendritic cells.

Host antimicrobial responses following the PRR-mediated signaling include proinflammatory reactions and immunogenic cell death[2]. When inflammation and other innate immune responses fail to combat the infection, infected cells opt to initiate diverse pathways that lead to immunogenic cell death[5]. These diverse forms of cell death play crucial roles in amplifying various downstream immune responses, restricting pathogen dissemination and eliminating infections[5]. Apoptosis, pyroptosis, necrosis and lysosomal cell death, representing the predominant and extensively investigated types of cell death triggered by pathogen infections[5], are discussed in this review.

Investigating the myriad manifestations of immunogenic cell death instigated by pathogenic infection is pivotal across several scientific domains. Such research provides pivotal insights into the pathophysiology of infectious diseases, facilitating the formulation of bespoke therapeutic and prophylactic approaches. Additionally, enhanced comprehension of immunogenic cell death mechanisms is instrumental in refining immunization strategies and therapeutic modalities to bolster endogenous immune defenses against pathogenic assaults. This review encapsulates the forefront of discoveries in delineating host receptors that recognize pathogen-associated molecular patterns, triggering the activation of inflammasomes and alternative signal transduction pathways, culminating in an array of immunogenic cell death phenotypes.

## 1. Host receptors is required for pathogen components recognition

The dynamic interaction between PAMPs and PRRs empowers the host to distinguish self-entities from foreign pathogens and to efficiently eradicate pathogenic invasions. Despite the immense diversity in microbial constitution, the host is nonetheless able to distinguish them through a small number of receptors using mechanisms that are strikingly similar yet significantly distinct[6]. Within the PRR family, members includes Toll-like receptors (TLRs), NOD-like receptors (NLRs), RIG-I-like receptors (RLRs), and C-type lectin receptors (CLRs)[1]. The activation of PRRs typically leads to the assembly of the inflammasome complex, a crucial sensor and mediator that subsequently triggers the activation of downstream inflammatory signalings[7]. The forefront of research focused on deciphering the complex interactions between hosts and pathogens, identifying novel PAMP-PRR interactions as well as the intricate mechanisms of recognition and the subsequent signaling cascades[2] (**Figure 1**).

### 1.1 Role of Toll-like receptors in immune surveillance

Among all PRRs, members of TLRs family have received the most attention from researchers in the past decades[8]. TLRs were initially identified due to their homology with the *Drosophila melanogaster* Toll protein, which acts as an immune guarder in the defense against fungal infections[9]. The observation that *Drosophila melanogaster* lacking Toll protein is susceptible to fungal infection contributed to the discovery of the importance of Toll protein in other species[10]. TLRs are expressed on cell membranes of diverse antigen presenting cells, including macrophages and dendritic cells. Though ubiquitously expressed in many scenarios, specific TLR expression can be inducible and exclusive to pathogen infections[11].

The TLR is composed of an extracellular domain containing the leucine-rich repeat (LRR) motif, and a Toll/interleukin-1 (IL-1) receptor (TIR) homology domain which facilitates intracytoplasmic signaling. Upon binding to pathogenic ligands, TLRs undergo oligomerization, triggering the onset of intracellular signal transmission[11]. In humans, ten TLRs have been identified to date, and these can be categorized into various subfamilies according to the PAMPs they recognize as well as their subcellular localizations. Whereas TLR1/ TLR2/ TLR4/ TLR5/ TLR6/ TLR10 are distributed on the surface of immune cells, TLR3/ TLR7/ TLR8/ TLR9 are mostly found on intracellular organelles, including endosomes and lysosomes[11]. TLRs on the cell surface can only recognize portions of bacteria, whereas TLRs on the endo-lysosomal membranes can bind to components of the host cell, thereby activating a wider variety of outcomes[12]. Among these TLRs, TLR5 and TLR4 are most extensively studied TLRs with regards to their involvements in sensing canonical components from bacterial infections[2].

TLR5 is able to detect flagellin, the subunit constituting the filament of bacterial flagella[13]. This structural element facilitates bacterial locomotion towards propitious environments and aids in the evasion of host immune defenses[14]. Upon TLR5 binding with flagellin, MyD88 is recruited to TRAF6, subsequently activating TAK1. This activation results in the phosphorylation of NF-κB and its subsequent translocation to the nucleus[15]. TLR5's detection of flagellin is an essential part of the innate immune responses to bacterial infections, especially those caused by motile bacteria such as *Salmonella Typhimurium* and *Pseudomonas aeruginosa*[15].

TLR4 is another crucial component of the host innate immunity and it's responsible for sensing Lipopolysaccharide (LPS) of gram-negative bacteria [16]. LPS is a complicated molecule composed of a lipid element that attaches it to the outer membrane and a polysaccharide portion that stretches outward[16]. On the one hand, LPS is recognized by TLR4 located on the cell membrane, further activating a MyD88-TAK1-NF-κB signaling cascade. On the other hand, LPS is detected by endosome-localized TLR4, activates TRIF-mediated IRF3/7 phosphorylation[17]. Both NF-κB and IRF3/7 activation promotes the expression of type I interferons (IFN-I) as well as pro-inflammatory cytokines, ultimately recruit and activate other immune cells to eradicate infections[17].

## 1.2 Recognition of nucleic acids by RIG-I like receptors

Beyond the recognition of outer membrane components of pathogens, host innate immune system also monitors the inner portions of these intruding bacteria that are released into host cells during infections[18], which includes their nucleic acids RNA and DNA[19]. RLRs are responsible for detecting exogenous RNA from both viral and bacterial infections, such as their double-stranded

RNA, 5'-triphosphate RNA, and short dsRNA[20]. For instance, RIG-I recognizes RNA fragments produced by *Listeria monocytogenes* and *Mycobacterium tuberculosis,* subsequently triggering the activation of the inflammasome and facilitating IFN-β production[21]. It is worth mentioning that dsRNAs are not exclusively recognized by RLRs, but also by other receptor types such as TLR3[22].

Exogenous cytosolic DNA derived from bacteria are also sensed by the host's receptors. These receptors include DNA-dependent activator of IFN-regulatory factors DAI, the cyclic GMP-AMP synthase (cGAS)-stimulator of interferon genes (STING) pathway, TLR9 and AIM2[1]. (1) DAI, the first identified cytoplasmic DNA sensor, exhibits a high affinity for DNA, thereby facilitating a robust immune response upon binding[23]. (2) Furthermore, the cGAS-STING pathway operates via a distinct mechanism[24]. Upon detecting the presence of foreign DNA, cGAS generates the secondary messenger cyclic GMP-AMP (cGAMP), which subsequently interacts with STING, instigating the production of IFN-I. (3) The TLR9 receptor specifically recognizes unmethylated CpG motifs, which are prevalent in bacterial DNA[25]. (4) AIM2 is adept at recognizing cytosolic dsDNA deriving from bacterial infections[26]. Upon dsDNA binding, AIM2 partners with the adaptor ASC and procaspase-1, leading to the activation of caspase-1 and the release of pro-inflammatory cytokines IL-1β and IL-18[26].

### 1.3 Intracellular sensor NOD-like receptors

NOD-like receptors (NLRs) are a class of intracellular PRRs that have piqued the interest of researchers because of their essential roles in recognizing bacteria that replicate intracellularly[27]. Being evolutionarily conserved proteins found in both vertebrates and invertebrates, NLRs are characterized with an N-terminal caspase recruitment domain (CARD), a central NOD domain and a C-terminal LRR domain. They are activated by a range of PAMPs, including flagellin, LPS, components derived from peptidoglycan from bacteria[28]. Members of NLR family includes Neuronal apoptosis inhibitory protein (NAIP), NOD1/2, NLRC3/4/5, NLRP1/3/6/12[29].

(1) NAIP5, a member of the NAIP family, was initially characterized for its capacity to detect flagellin originating from *L. pneumophila*[30]. Macrophages derived from mice with multiple polymorphisms in the *Naip5* gene display increased susceptibility to *L. pneumophila* infections[31]. (2) NOD1 and NOD2 are specialized in the detection of numerous components from pathogenic bacteria[32]. NOD1 senses diaminopimelic acid from Gram-negative bacteria, such as *Shigella flexneri* and *Pseudomonas aeruginosa* [33]. NOD2 recognizes muramyl dipeptide found in both Gram-positive and -negative bacterial species, including *Mycobacterium tuberculosis* and *Listeria monocytogenes*[34]. (3) Besides, the lethal toxin produced by *Bacillus anthracis* is recognized by NLRP1b, culminating in its cleavage and subsequent activation of inflammasomes[35]. (4) Additional ligands detected by NLRs encompass bacterial peptidoglycan and diminished level of cytosolic ATP[27].

Upon activation, NLRs undergo the conformational change, allowing them to oligomerize via CARD-CARD interactions. This connection further recruits an adaptor protein and a pro-caspase, assembling the multimeric protein complex, inflammasome, whose activation culminates in the secretion of pro-inflammatory cytokines IL-1β and IL-18 or the induction of pyroptotic cell death[32]. Collectively, these receptor systems provide a comprehensive defense mechanism, allowing the host to detect and respond to invading pathogens effectively.

**1.4 Mechanism of inflammasome activation**

The inflammasome is a multi-protein complex assembled upon detection of PAMPs by PRRs, driving host defense mechanisms. It is typically constituted of a nucleotide-binding domain and LRR-containing protein, the adaptor protein ASC and pro-caspases[36]. Two signals are required for the inflammasome to be activated. The first signal is priming, which involves the recognition of PAMPs by membrane bound or cytoplasmic PRRs. This causes the upregulation of pro-IL-1 and pro-IL-18[7]. The second signal is provided by the inflammasome complex itself, which triggers the oligomerization of NLRs and the recruitment of ASC and pro-caspase-1[7]. This complex then undergoes a conformational change that results in the activation of caspase-1 and the processing of pro-IL-1β and pro-IL-18 into their mature biologically active forms[37,38]. The coordinated interplay of these sequential processes ultimately facilitates the activation of immune responses and efficient elimination of invasive pathogens.

Many varieties of inflammasomes have been identified so far; each is activated by specific stimuli and composed of diverse members of NLR family and caspase effectors[36]. (1) The NLRP3 inflammasome, comprising the receptor NLRP3, the adaptor protein ASC, and pro-caspase-1, represents the most extensively analyzed inflammasome assembly. Its activation can be triggered by a broad spectrum of PAMPs, including bacterial RNA, DNA viruses[39], and fungi[40]. The priming for canonical NLRP3 inflammasome activation is mediated by caspase-8 while non-canonical NLRP3 inflammasome activation requires the binding of caspase-11 and cytosolic LPS[40].

(2) Another important inflammasome is the NAIP-NLRC4 inflammasome, which can be activated by cytosolic bacterial flagellin and needle proteins, inner rod proteins of the type III secretion system (T3SS) of pathogenic bacteria[41]. Interestingly, proteins involved in cellular metabolic pathways, which are subject to caspase-1-mediated cleavage during infection with *Salmonella Typhimurium*, also instigate the activation of the NLRC4 inflammasome[42].

(3) AIM2 inflammasome plays a critical role in sensing cytosolic dsDNA from invading bacteria[26]. Structural analyses reveal that upon binding to dsDNA, AIM2 oligomerizes and recruits ASC, which then recruits pro-caspase-1 to form the inflammasome complex[43]. Certain bacteria encode effectors that allow them to escape detection by the AIM2 inflammasome. *L. pneumophila*, for instance, employs SdhA to maintain the integrity of the bacterial replicative vacuole, preventing the leakage of DNA into the cytoplasm[44].

(4) Lastly, recent investigations have unveiled a unique mechanism employed by the Pyrin inflammasome to identify bacterial infections[45]. Recent investigations have unveiled a unique mechanism employed by the Pyrin inflammasome to identify bacterial infections[46]. In this context, Pyrin recognizes the bacterial-induced modifications of Rho GTPases, such as the glycosylation by *Clostridium difficile* toxins TcdA/B[47] and the mono ADP-ribosylation by the C3 exoenzyme from *Clostridium botulinum*[48]. Additional bacterial-induced modifications, including adenylation and deamidation, also serve as initiating stimuli for the activation of the Pyrin inflammasome[36].

## 2. Immunogenic cell death triggered by the pathogen infections

One effective strategy of host immune defense in response to the infection of pathogens is induction of cell death[5], an event that will eliminate the niche for pathogen propagation. Investigation of host responses including cell death triggered by pathogens has led to the identification of pattern recognition receptors and novel immune mechanisms[1]. The dynamic interaction between pathogen associated molecular patterns and host germline-encoded pattern-recognition receptors empowers the host to distinguish self-entities from foreign pathogens and to efficiently eradicate pathogens. Despite the immense diversity in microbial constitution, the host is nonetheless able to distinguish them through a small number of receptors using mechanisms that are strikingly similar yet significantly distinct[6].

**2.1 Apoptosis**

Apoptosis is a highly regulated form of cell death commonly observed in normal development and during pathogen infections[49]. This process is a multifaceted orchestration involving a series of messengers and enzymes, including members of the caspase family and mitochondrial-associated proteins without discharging cellular components into the extracellular lilieu[50]. Hence, apoptosis is considered to proceed without eliciting an inflammatory response. Therefore, it is considered immunologically silent[5], distinct from the inflammatory cell death pyroptosis[51]. Characteristics of infected cells undergoing apoptosis include DNA fragmentation, nuclear condensation[52], cytoplasmic blebbing, cell shrinkage and the formation of apoptotic bodies[53]. Induction of apoptosis facilitates the removal of infected cells, thus preventing the spread of pathogens into deep tissues.

The induction of apoptosis by infection occurs through multiple distinct pathways: (1) Caspases are proteases that are activated in a cascadic manner upon specific apoptotic stimuli, such as LPS and Fas ligand (FasL)[54]. More precisely, the interaction between FasL derived from *Helicobacter pylori* and its cognate Fas receptor (CD95)[55] and tumor necrosis factor receptor 1 (TNFR1), triggers the assembly of death-inducing signaling complex (DISC), resulting in the activation of caspases[56]. (2) Infection-induced apoptosis can also occur through the impairment of mitochondrial integrity, accompanied by the release of pro-apoptotic factors. For example, the toxin EspC secreted by *Mycobacterium tuberculosis* triggers the permeabilization of the outer mitochondrial membrane (MOMP), allowing the release of cytochrome c, Calcium ions ($Ca^{2+}$) and other apoptogenic factors into the cytoplasm[57]. Cytochrome c then activates caspase-9, which initiates the intrinsic apoptotic pathway[58]. (3) Recent studies show that infections by *L. pneumophila* lead to extensive apoptosis in specialized phagocytes, such as dendritic cells[59]. From a molecular perspective, infections by these pathogens tips the balance between the pro-apoptotic and anti-apoptotic constituents of the Bcl-2 protein family, leading to the initiation of the MOMP and subsequent activation of caspase-3-mediated apoptosis[59]. Intriguingly, infections of *L. pneumophila* in permissive macrophage did not exhibit obvious apoptosis[60], suggesting that *L. pneumophila* possesses mechanisms to prevent infected macrophages from apoptotic cell death. This hypothesis gained experimental support when it was observed that infections by *L. pneumophila* strains lacking *sdhA* or *sidF* elicited enhanced the induction of apoptosis in macrophages[61]. Whereas SidF appears to function by inhibiting the activity of a pro-death member of the Bcl2 protein family[62], ShdA functions by maintaining the integrity of the bacterial phagosome[63] (**Figure 2**).

In light of the immune-defense functions of apoptosis, pathogens have evolved a wide array of strategies to counteract and inhibit apoptosis, thereby ensuing their successful replication within host cells. (1) Some bacterial species synthesize proteins that specifically engage and proteolytically cleave vital elements of the host's apoptotic pathways. An illustration of this is the AIP56 toxin secreted by *Photobacterium damselae piscicida*, which catalyzes the cleavage of NF-κB p65, consequently inhibiting the NF-κB-dependent transcription of pro-inflammatory genes[64]. (2) Certain bacterial pathogens prevent apoptosis by modulating autophagy[65,66]. For example, *Salmonella* Typhimurium can switch the fate of host cells by triggering autophagy and preventing infected cells from undergoing apoptosis[67]. This is achieved by the leakage of amino acids from the pores formed by its T3SS1, which activates acute starvation stress, triggering eIF2α/ATF4-meadiated autophagy pathway[68]. (3) Many bacteria are capable of inhibiting the initiation of apoptosis by inducing the transcription of anti-apoptotic genes. For example, *L. pneumophila* infection induces the activation of the MAP kinase pathway in a Dot/Icm dependent manner, resulting in increased expression of anti-apoptotic proteins[69]. (4) Effector proteins secreted by some bacteria directly hijack constitutes of apoptotic pathways. As an illustration, *L. pneumophila* effector SidF selectively antagonizes the activities of two pro-apoptotic Bcl2 members, thereby impedes apoptosis of infected cells[62]. In summary, bacterial pathogens employs complex approaches with multiple effectors to manipulate host apoptosis pathways to counteract elimination caused by cell death. These balancing acts between apoptosis induction and inhibition highlights the evolutionary mechanisms pathogens adopt to thrive within hosts.

## 2.2 Pyroptosis

Pyroptosis is a highly inflammatory form of programmed cell death involved in the host defenses against microbial infections[70]. It is initiated by the activation of the inflammasome which senses PAMPs or DAMPs derived from the invading pathogens or damaged host cells[71]. Activation of inflammasomes leads to the activation of caspases which cleave cytokine precursors and/or members of Gasdermin family[72] to release the N-terminal portion of these proteins to form pores in the plasma membrane, leading to cell swelling, osmotic imbalances and the leakage of cellular contents[73]. These ultimately result in the lysis of infected cells and the release of inflammatory signals.

Pyroptosis is mediated through various intricate mechanisms and pathways[70]. (1) A selection of inflammasomes, including Nucleotide-binding Oligomerization Domain (NOD)-like receptor (NLR) family Pyrin domain-containing protein 3 (NLRP3), Absent In Melanoma 2 (AIM2), Pyrin, and NLR family CARD domain-containing protein 4 (NLRC4), orchestrate the activation of caspase-1[26]. For example, upon recognition of flagellin from *L. pneumophila* by NAIP5[74], the NLRC4 inflammasome recruits and activates caspase-1, which then cleaves GSDMD. (2) Alternatively, caspase 4/5/11 directly senses cytosolic bacterial LPS and activates itself to cleave and activate GSDMD[75]. In both scenarios, pro-IL-1β and pro-IL-18 are cleaved by caspase-1, leading to the release of mature cytokines and intracellular DAMPs, thereby amplifying the inflammatory responses against the pathogen invasions[71]. (3) Similarly, Gasdermin E (GSDME) undergoes specific cleavage by caspase 3, with the resultant N-terminal fragment instigating pyroptosis as a countermeasure against bacterial infection[76]. Interestingly, the caspase-3-GSDME axis may also be activated by granzyme B (GZMB) within lung alveolar epithelial cells infected by H7N9 virus, resulting in an overwhelming cytokine response and pyroptosis[77]. (4) Furthermore, cytotoxic T lymphocytes (CTLs) and natural killer (NK) cells possess the capability to secrete serine proteases granzymes, targeting infected cells or cancer cells[78]. Granzyme A (GZMA) released from these cytotoxic lymphocytes cleaves Gasdermin B (GSDMB) within targeted cells. The N-terminal domain of GSDMB forms pores on membranes, leading to pyroptosis[79]. (5) In a separate mechanism, the AIM2 inflammasome detects cytosolic dsDNA signatures from invading bacteria[26], binding to dsDNA induces AIM2 oligomerization, which in turn recruits the adaptor

Apoptosis-associated Speck-like protein containing a CARD (ASC)[80]. This complex then recruits pro-caspase-1, paving the way for inflammasome assembly and the induction of pyroptosis[43].

Intriguingly, the proteolytically cleaved form of GSDMD possesses the ability to directly lyse bacteria by assembling pores in the bacterial cell membrane[81]. More specifically, this happens when its N-terminal fragment binds with cardiolipin, a phospholipid localized in the cell membranes of bacterial species such as *Staphylococcus aureus* and *Bacillus megaterium*[82].

To counter the damage caused by pyroptosis, bacterial pathogens have evolved effective strategies to inhibit the activation of pyroptosis within infected host cells[83]. (1) Among these, *Yersinia pestis* capitalizes on the functionalities of its effectors YopK and YopM. While YopK inhibits the recognition of its T3SS by the NLRC4 inflammasome[84], YopM inhibits the activation of the Pyrin inflammasome[85]. This dual action ultimately hinders caspase-1 activation, effectively suppressing the initiation of pyroptosis. (2) *L. pneumophila*, for instance, employs its effector SdhA to maintain the structural integrity of its replicative vacuole, thus preventing the leakage of DNA into host cytoplasm[44], which will avoid pyroptosis caused by AIM2 activation[44] and by type I IFN induction[86]. (3) Furthermore, *Shigella flexneri*, implicated in bacillary dysentery, inhibits LPS-induced pyroptosis via its effector OspC3. OspC3 catalyzes arginine ADP-riboxanation on caspase-4/-11, halting the proteolytic processing of GSDMD and subsequent pyroptosis[87]. Similar to *Shigella* OspC3, the effector CopC, secreted by *Chromobacterium violaceum*, also possesses the ADP-riboxanase activity. Once specific interacting with host calmodulin (CaM), CopC mediates arginine ADP-riboxanation of apoptotic caspases encompassing caspase-7/-8/-9[88]. Collectively, bacterial pathogens deploy a myriad of effectors to intricately modulate and impede host pyroptosis, highlighting the perpetual evolutionary interplay between microbial virulence and host immunity.

## 2.3 Necrosis

Necrosis is another type of cell death induced by pathogen infections[89] characterized by rupture of the plasma membrane, nuclear swelling and release of cellular contents into the extracellular space[90], resulting in inflammation independent of caspases[91]. During pathogen infections, several mechanisms can lead to necrosis:

(1) Certain pathogens produce toxins or enzymes that directly damage host cells and lead to necrosis. For example, the alpha-toxin of *Staphylococcus aureus* causes the formation of pores in the plasma membrane, resulting in cell swelling and lysis[92]. (2) Besides, the replication of bacteria in host cells leads to altered homeostasis and the accumulation of misfolded proteins, resulting in endoplasmic reticulum (ER) stress and subsequent necrosis[93]. (3) Moreover, pathogen infections can induce necrosis via the production of reactive oxygen species (ROS)[94]. For example, the uracil released by the *Bacillus thuringiensis* promotes mitochondrial dysfunction[95] and the activation of NADPH oxidases, which then leads to the production of ROS[96]. ROS subsequently induces oxidative stress and damages cellular components, resulting in necrosis[97]. (4) Lastly, cellular $Ca^{2+}$ dysregulation during infections also triggers necrosis in infected cells, the disruption of $Ca^{2+}$ homeostasis lead to an influx of $Ca^{2+}$ into the cytoplasm[98]. Excessive cytoplasmic $Ca^{2+}$ levels can activate various enzymes that perturb cellular processes, and ultimately lead to necrotic cell death.

The consequences of pathogen-induced necrosis are usually detrimental to both the infecting organisms and the hosts[99]. Cellular contents released upon cellular damage or inflammation can lead to the activation of auto-immune responses and the amplification of inflammation[100], thereby exacerbating tissue destruction as well as broader systemic effects.

**2.4 Lysosomal cell death**

Lysosomes are membrane-bound organelles containing various hydrolytic enzymes involved in intracellular degradation and cell recycling[101]. During bacterial infections, lysosomes can exhibit both beneficial and detrimental effects on infected cells depending on the magnitude of lysosomal perturbations[91]. Concurrently, lysosomes contribute to host defenses by fusing with phagosomes to degrade the engulfed bacteria[102]. This process aids in the eradication of intracellular pathogens and prompts immune reactions. However, microbial infections can compromise lysosomal membrane integrity, a phenomenon referred to as lysosomal membrane permeabilization (LMP)[103]. The occurrence of LMP is accompanied with release of cathepsins from the lysosomal lumen, which leads to the cleavage of Bid to generate tBid[104]. tBid then forms pores in mitochondria to trigger the release of cytochrome c (Cyto c), which activates the classic apoptotic pathway that ultimately leads to caspase-3 activation and cell death[105,106] (**Figure 3**).

Bacterial infections can trigger lysosomal cell death via multiple mechanisms[107]: (1) Toxins or pore-forming proteins secreted by some bacteria directly target lysosomal membranes, leading to LMP, such as the nigericin from *Streptomyces hygroscopicus* and the pyocyanin from *Pseudomonas aeruginosa*[106]. (2) In some context, sensing of bacterial ligands by host receptors leads to lysosomal destabilization. The effector VepA secreted by *Vibrio parahaemolyticus* interacts with $H^+$-ATPase and such binding disrupts the integrity of lysosomal membranes and induces LMP[108]. (3) In addition, the accumulation of ROS produced during intracellular bacterial infections cause oxidative stress and damages to lysosomal membrane proteins. For instance, infections by *Shigella* or *Chlamydia* are detected by NLR Family Member X1 (NLRX1), which localizes to mitochondria and induces the production of ROS[109], resulting in subsequent lysosomal cell death. (4) An earlier study from our lab discovered that infections by *L. pneumophila* strains harboring wild-type *rpsL* such as Lp02*rpsL*$_{WT}$ induce extensive lysosome damage and apoptosis in mouse bone marrow-derived macrophages (BMDMs), resulting in the termination of bacterial replication[103]. Although the mechanism of this unique infection-induced cell death remains unknow, lysosomes appear to be involved[110]. Cellular events upstream of lysosomal membrane permeabilization awaits further investigation[111].

**Discussion**

The intricate interplay between host immune defenses and pathogen evasion strategies, as outlined in this review, emphasizes the pivotal role of immunogenic cell death in controlling pathogen infections. Central to this complex dance is the activation of host receptor-mediated signaling pathways as well as downstream cell death[112], enabling the host to detect and respond effectively to invading pathogens[56]. Such immune responses often include the activation of the inflammasome complex, which further augments cell death and amplifies inflammation[113]. The distinct mechanisms of immunogenic cell death not only facilitate the elimination of infected cells but also significantly contribute to training subsequent adaptive immune responses[114]. Yet, pathogens are extremely smart in that they have evolved sophisticated mechanisms to subvert these host defenses, either by directly disrupting the function of key factors in immunogenic cell death pathways or by interfere the activity of components critical for inflammation[82]. The ultimate aim is to manipulate host cells to favor pathogen survival and replication.

Immunogenic cell death in response to pathogen invading is a double-edged sword. On one hand, cell death represents a defense mechanism that efficiently clears infected cells to limit pathogen spread[91]. On the other hand, excess inflammation cause damage in mucosal barrier and tissues, an opportunity that can be exploited by infectious microbes[70]. In addition, certain pathogens have developed strategies to inhibit cell death, thereby enhancing their survival within the host. This tug-of-war between cell death induction and inhibition mirrors the evolutionary battle between host defenses and microbial virulence[115]. Although apoptosis is the type of cell death most extensively studied, the roles of pyroptosis and necrosis in immune responses demonstrate the host's ability to utilize diverse cell death mechanisms to combat infections[116]. In particular, pyroptosis, which releases inflammatory signals, serves as a critical defense strategy by recruiting additional adaptive immune cells.

New insights into these dynamic host-pathogen interactions are unfolding, shedding light on immune defense mechanisms and pathogen evasion tactics[28]. It becomes evident that the host's ability to initiate various different forms of cell death is a critical factor in the outcome of infections. The ongoing discovery of novel receptors, signaling pathways, and microbial evasion tactics deepens our understanding of the immune system's complexity and adaptability[53]. Future research should delve deeper into these complex molecular mechanisms, potentially revealing novel therapeutic targets for emerging infectious diseases[117]. Understanding how pathogens modulate immunogenic cell death pathways might lead to innovative immune response strategies, improving the efficacy of treatments against various infectious agents[75]. In summary, further studying the diverse mechanisms of immunogenic cell death triggered by pathogen infections is a multidisciplinary endeavor that has far-reaching implications for medicine, public health, and our understanding of human biology.

**Acknowledgments**

We thank Dr. Rui Cao of Cornell University for critically reviewing the manuscript and for offering significant insights.

**Declaration of interests**

The authors declare no competing interests.

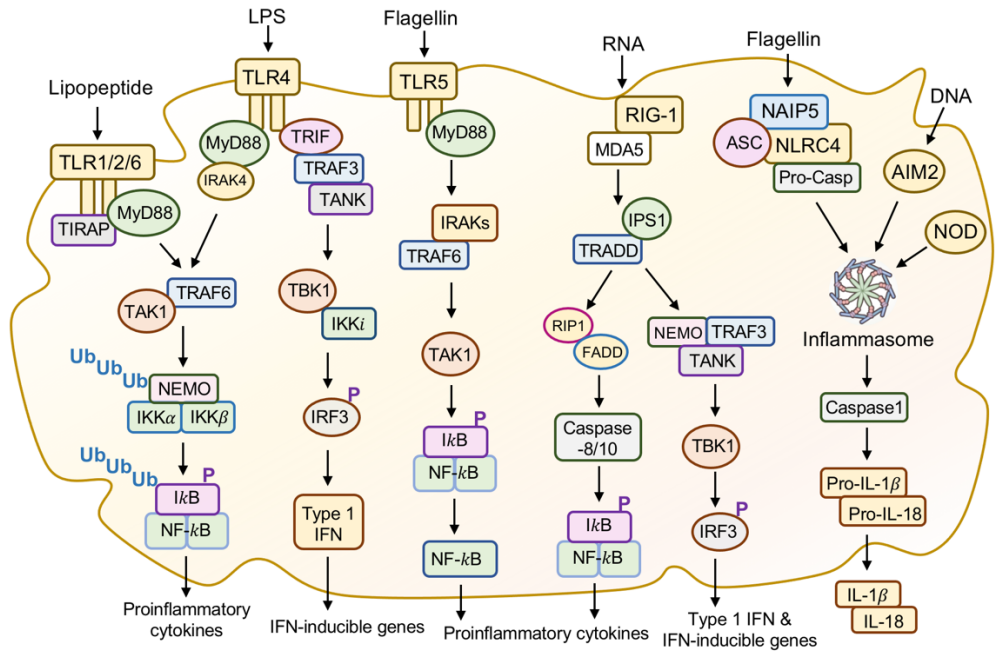

**Figure 1: Signaling Pathways of Innate Immune Responses**

The schematic graph illustrating the complex network of signaling pathways involved in the innate immune response, highlighting the role of various pattern recognition receptors (PRRs) such as TLRs (Toll-like receptors), RIG-1 (Retinoic acid-inducible gene I), and NOD-like receptors (NLRs) in detecting pathogen-associated molecular patterns (PAMPs) like lipopeptides, flagellin, RNA, and DNA. Upon activation, these receptors initiate a cascade of interactions involving adaptor proteins (e.g., MyD88, TRIF, TRADD), kinases (e.g., IRAKs, TAK1, TBK1), and other molecules (e.g., NEMO, IKK complex, IRF3) that ultimately lead to the transcriptional activation of NF-κB and IRF3. This results in the production of proinflammatory cytokines and type I interferons (IFN), which are crucial for the inflammatory response and the establishment of an antiviral state. The assembly of the inflammasome, a multiprotein oligomer that activates caspase-1, is depicted, leading to the processing and secretion of proinflammatory cytokines IL-1β and IL-18. Additionally, the figure maps the ubiquitination (Ub) and phosphorylation (P) events that regulate these pathways, emphasizing the tight regulation of signaling required for an appropriate immune response.

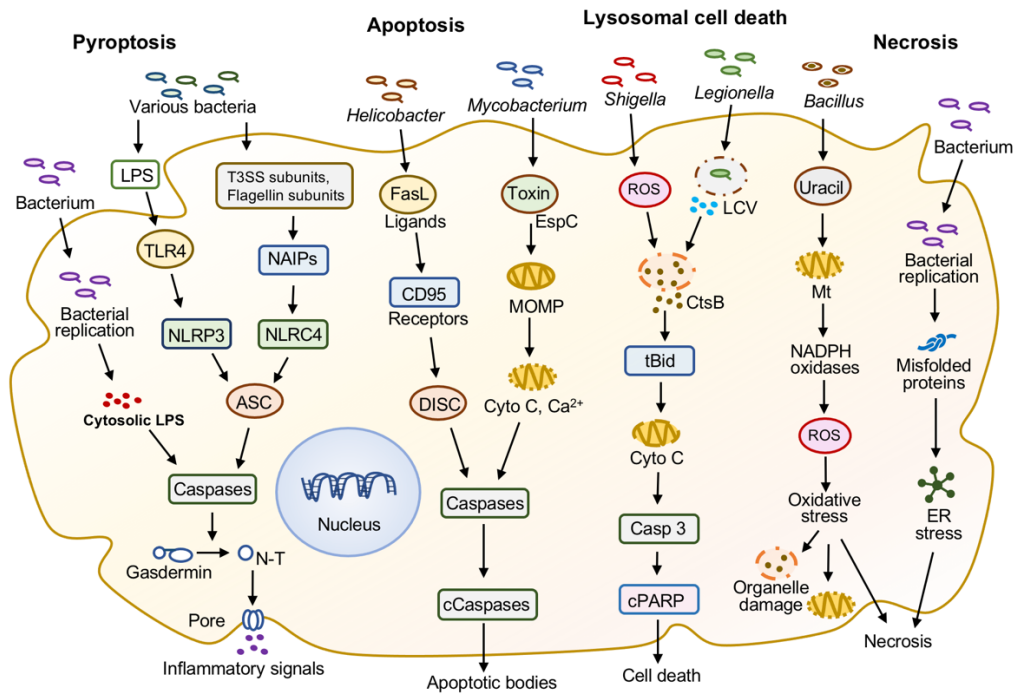

**Figure 2: Mechanisms of Pathogen-Induced Immunogenic Cell Death**

The mechanisms of cell death initiated by various bacterial pathogens is divided into four main pathways: pyroptosis, apoptosis, lysosomal cell death, and necrosis. **(1)** Pyroptosis is triggered by bacterial components like LPS, which are recognized by TLR4, leading to the activation of NLRP3 or NLRC4 inflammasomes. **(2)** Apoptosis is depicted as being induced by several bacterial strategies, including the activation of death receptors. This leads to the formation of the death-inducing signaling complex (DISC), the release of cytochrome c from mitochondria, and the activation of caspases that result in apoptotic body formation and cell death. **(3)** Lysosomal cell death is shown to be initiated by reactive oxygen species (ROS) and involves lysosomal membrane permeabilization (LMP), leading to the release of cathepsins, which contribute to the activation of Bid and subsequent cytochrome c release and caspase activation. **(4)** Necrosis is illustrated as being caused by factors such as uracil from bacteria that lead to mitochondrial (Mt) dysfunction and oxidative stress through NADPH oxidases. This results in reactive oxygen species (ROS) production, organelle damage, and ultimately, necrotic cell death characterized by a loss of membrane integrity and uncontrolled release of cell contents.

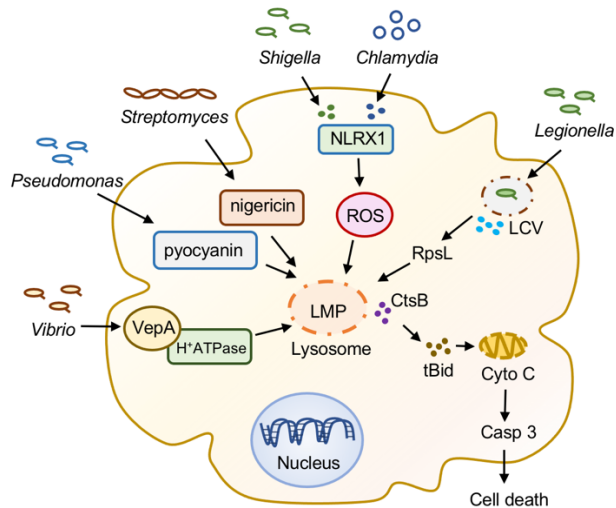

**Figure 3: Approaches employed by Pathogens in Lysosomal Cell Death**

Bacteria induce cell death through lysosomal damage and the subsequent cellular pathways that lead to apoptosis. It depicts various bacterial species releasing toxins and other factors that converge on the lysosome, causing its destabilization. *Pseudomonas* produces pyocyanin, which contributes to the generation of reactive oxygen species (ROS) within the cell. *Vibrio* species release VepA, which inhibits the $H^+$ ATPase function. *Streptomyces* contributes nigericin as part of its pathogenicity mechanism. Pathogens like *Shigella* and *Chlamydia* are shown to manipulate host cell functions, involving the recognition of cellular damage by the NLRX1 receptor, which then triggers the production of ROS. The increase in ROS leads to lysosomal membrane permeabilization (LMP), releasing cathepsins (CtsB) into the cytosol. *Legionella* RpsL, a protein associated with antibiotic resistance, contributing to the cellular stress and lysosomal damage. The released cathepsins activate Bid, a pro-apoptotic molecule, which in turn leads to the release of cytochrome c (Cyto C) from the mitochondria. Cytochrome c then initiates the activation of caspase 3 (Casp 3), culminating in cell death.